\documentclass[preprintnumbers]{iopart}
\usepackage{graphicx}
\usepackage{amssymb}

\usepackage{color}

\def\qaq{\quad\textrm{and}\quad}
\def\O{\mathcal{O}}
\def\F{\mathcal{F}}
\def\P{\mathcal{P}}

\def\cross{\times}
\def\hplus{{h_+}}
\def\hcross{{h_\cross}}

\def\vn{\vec{n}}
\def\vr{\vec{r}}
\def\vb{\vec{\beta}}
\def\LL{\Lambda}
\def\const{\mathrm{const.}}

\def\ML{{\mathrm{ML}}}
\def\sig{{\mathrm{s}}}

\def\spin{{\mathrm{spin}}}
\def\orb{{\mathrm{orb}}}

\def\vA{\vec{A}}
\def\KA{K_A}

\def\lat{\lambda}
\def\RE{R_\oplus}

\def\ez{\hat{z}}
\def\ew{\hat{\varpi}}

\def\Om{\Omega}
\def\sinc{\textrm{sinc}}
\def\round{\textrm{round}}

\def\Hz{\textrm{Hz}}

\def\implies{\Longrightarrow}
\def\th{{\mathrm{th}}}

\begin{document}

\title[Global parameter-space correlations of coherent pulsar-searches]{
  Global parameter-space correlations of coherent searches for
  continuous gravitational waves.}
\author{Reinhard Prix and Yousuke Itoh}
\address{
Max Planck Institut f\"ur  
Gravitationsphysik, Albert Einstein Institut, \\ 
Am M\"uhlenberg 1, Golm 14476, Germany
}

\ead{Reinhard.Prix@aei.mpg.de}

\begin{abstract}
The space of phase-parameters (sky-position, frequency, spindowns) of
a coherent matched-filtering search for continuous gravitational
waves from isolated neutron stars shows strong global correlations
(``circles in the sky'').    
In the local limit this can be analysed in terms of a parameter-space
metric, but the global properties are less well studied.
In this work we report on our recent progress in understanding these
global correlations analytically for short to intermediate (less than
a month, say) observation times and neglecting spindowns.
The \emph{location} of these correlation-circles in parameter-space is
found to be determined mostly by the orbital velocity of the
earth, while the spin-motion of the detector and the antenna-patterns
only contribute significantly to the amplitude of the detection
statistic along these circles.
\end{abstract}

\pacs{04.80.Nn, 95.75.-z\\
AEI publication number: AEI-2005-077}

%% 04.80.Nn     Gravitational wave detectors and experiments 
%% 95.75.-z     Observation and data reduction techniques; computer modeling and simulation

\section{Introduction}

Continuous gravitational waves from, for example, rotating
neutron stars with non-axisymmetric deformations such as mountains
or oscillation-modes, are one of the primary targets of
current-generation detectors of both interferometric- (e.g.
see~\cite{lsc:_settin_upper_limit,lsc05:_limit_ligo_selected_pulsars})
as well as bar-detector design~\cite{astone03:_all_sky}.

In this paper we restrict ourselves to gravitational waves emitted
from isolated neutron stars with negligible proper motion, for which the
signal can be assumed to be a nearly monochromatic sinusoidal (with
slowly decreasing frequency) in the solar-system barycenter frame (SSB).
The corresponding signal received at the detector will be
Doppler-modulated by the spin of the earth and its orbital motion
around the sun.  
The phase of the received signal therefore depends not only on its
intrinsic frequency $f$ and spindowns $f^{(k)}\equiv d^k f/dt^k$,
but also on the sky-position (denoted by $\alpha$ and $\delta$ for
right ascension and declination) of the source and on the detector
location.  
In addition there is a time-dependent amplitude modulation of
the signal depending on its polarisation angle $\psi$ and amplitudes
$\hplus$ and $\hcross$. However, as shown in~\cite{jks98:_data}, one
can eliminate these ``amplitude parameters'' together with the initial
phase of the signal $\Phi_0$ by analytically maximising the detection
statistic over these parameters. The resulting reduced parameter-space therefore
only consists of the ``phase parameters'' $\P=\{\alpha, \delta, f, f^{(k)}\}$, 
and the corresponding partially maximised detection statistic is
usually referred to as the ``$\F$-statistic''. 

Given a stretch of data $x(t)$ from the detector, an untargeted
search (as opposed to a search for known pulsars) consists of
calculating the detection statistic $\F$ over the parameter space (by
sampling individual points) and determining the location of
``candidates'' that cross a predetermined threshold.  
For a given signal $\P_\sig=\{\alpha_\sig, \delta_\sig, f_\sig,
f^{(k)}_\sig\}$, the detection statistic $\F$ in a search-point $\P$
generally depends on the parameter-mismatch
$\Delta\P\equiv\P_\sig-\P$, i.e., $\F = \F(\P_\sig, \Delta \P)$
(strictly speaking, $\F$ also depends on the amplitude parameters of the signal).  
Neglecting noise, one can expect $\F$ to be maximal for the perfectly
matched search-point $\Delta\P=0$. In the local neighbourhood 
of this point, the loss 
$\Delta\F \equiv \F(\P_\sig, 0)-\F(\P_\sig,\Delta\P)$ in
detection statistic can be quantified in terms of a parameter-space
metric as discussed in~\cite{owen96:_search_templates,brady98:_search_ligo_periodic}.
Namely, neglecting effects of the amplitude-parameters of the
signal, this metric can be defined as
$\Delta\F(\P_\sig,\Delta\P) / \F(\P_\sig,0) = \sum_{i,j} g_{i
    j}(\P_\sig)\,\Delta\P^i\,\Delta\P^j + \O(\Delta\P^3)$.

In the case of isolated neutron stars, we find that this
metric is \emph{highly} anisotropic in terms of the above
parameter-space variables (in particular with respect to
sky-position), as will also be seen in \sref{sec:comp-fully-numer}. 
Furthermore, the global behaviour of $\F$ differs dramatically from the
local quadratic decrease as a function of parameter-mismatch
$\Delta\P$.  This can be seen, for example, in \fref{fig:circlesF} 
for an all-sky search at fixed frequency of an injected signal. 
For each target search-frequency $f$ within the Doppler-window of the
signal frequency $f_\sig$, one finds a very thin circular band on the
sky along which $\F$ is of comparable magnitude to its maximum, while
it drops sharply to zero in the directions orthogonal to this
``circle''. In the local limit this feature has already been pointed
out in~\cite{krishnan04:_hough}. 

In this work we present an approximate analytic description of these
``circles in the sky'', tracing their physical origin to the
Doppler-shift due to the orbital motion around the sun.
The Doppler-shift induced by the spin-motion of the earth as
well as the amplitude modulations due to the rotating antenna-pattern
are found to give only very small corrections to the \emph{structure}
of these patterns in parameter-space (i.e., the location of the
circles), while they do contribute significantly to the
\emph{amplitude} of the $\F$-statistic along these circles. 

\section{Matched filtering of continuous signals}

The gravitational-wave strain $h(t)$ at the detector can be written in 
the form 
\begin{equation}
  \label{eq:1}
  h(t) = F_+(t)\, \hplus(t) + F_\cross(t)\, \hcross(t)\,,
\end{equation}
where $F_{+,\cross}$ are the antenna-pattern functions, and 
$h_{+,\cross}$ are the two polarisation components of the
gravitational wave. For a continuous pulsar-signal we can write
\begin{equation}
  \label{eq:2}
  \hplus(t) = A_+ \sin \left( \Phi(t) + \Phi_0 \right)\,,\quad
  \hcross(t) = A_\cross \cos \left(\Phi(t) + \Phi_0 \right)\,.
\end{equation}
The phase $\Phi(t)$ of the signal is assumed to be of the form
\begin{equation}
  \label{eq:3}
  \Phi(t; f^{(k)}, \vn) = 2\pi \sum_{k=0}^s {f^{(k)} \over (k+1)!} \,\tau^{k+1}(t,\vn)\,,
\end{equation}
where $s$ is the number of spindown parameters, and $\tau$ is the
arrival time in the solar-system barycenter (SSB) of a wave-front from
direction $\vn = (\cos\delta \cos\alpha, \cos\delta \sin\alpha, \sin\delta)$
arriving at the detector at time $t$, i.e., 
\begin{equation}
  \label{eq:4}
  \tau(t,\vn) \equiv t + { \vr(t)\cdot\vn \over c}\,,
\end{equation}
with $\vr(t)$ denoting the vector from the SSB to the detector. As
shown in~\cite{jks98:_data}, the full phase-model \eref{eq:3} can be
simplified by an approximated model without significant loss in 
detection statistic, namely
\begin{equation}
  \label{eq:5}
  \Phi(t) \approx 2\pi\left[ \sum_{k=0} {f^{(k)} \over (k+1)!} t^{k+1} +
  {\vr(t)\cdot\vn \over c} \sum_{k=0} {f^{(k)} \over k!} t^k \right]\,,
\end{equation}
which will be dominated by the first term, i.e., $\Phi(t)\approx 2\pi f t$. 
In the case of stationary Gaussian noise, the likelihood ratio $\LL$ (which
is the optimal detection statistic in the sense of Neyman-Pearson) is
given  by 
\begin{equation}
  \label{eq:6}
  \ln \LL = {T \over S_h} \left[ (x||h) - {1\over2} (h||h) \right]\,,
\end{equation}
where $S_h$ is the power spectral density of the noise, and the scalar
product $(x||y)$ is defined as  
\begin{equation}
  \label{eq:7}
  (x||y) \equiv {2\over T} \int_0^T x(t)\,y(t)\,d t\,.
\end{equation}

\section{Simplified matched-filtering statistic: neglecting amplitude modulations}

In the following we make the simplifying assumption that we can neglect
the amplitude-modulation caused by the antenna-pattern, so we assume
$F_{+,\cross}\approx \const$, and using 
\eref{eq:1} and \eref{eq:2} this results in the simplified strain model
\begin{equation}
  \label{eq:13}
  h(t) = A_1 \,\cos \Phi(t) + A_2 \,\sin \Phi(t)\,.
\end{equation}
Maximising the log-likelihood function \eref{eq:6} over the two
unknown amplitudes $A_{1,2}$, we obtain 
\begin{equation}
  \label{eq:14}
  \ln \LL_\ML 
%%={T\over 2 S_h}\left[ (x||\cos\Phi)^2 + (x||\sin\Phi)^2\right] 
  = {T\over 2 S_h}\left| \left(x||e^{-i\Phi} \right) \right|^2 
\equiv {T\over 2 S_h} |X|^2\,,
\end{equation}
defining the matched-filtering amplitude $X$.
In the following we assume the data contains only a unit-amplitude
pulsar-signal of the form   
\begin{equation}
  \label{eq:17}
  x(t) = \Re[s(t)],\quad\textrm{with}\quad
  s(t) = \exp\left[ i \,\Phi\left(t; f_\sig^{(k)}, \vn_\sig\right)\right]\,,
\end{equation}
characterised by the signal phase-parameters $f_\sig^{(k)}$ and $\vn_\sig$.
The matched-filtering amplitude $X$, defined in \eref{eq:14}, can therefore
be expressed as
\begin{equation}
  \label{eq:16}
  X = \left(x||e^{-i\Phi}\right) \approx {1\over T}\int s(t)\, e^{-i\Phi(t)}\, d t
  = {1\over T}\int e^{i \Delta\Phi(t)}\,d t\,,
\end{equation}
where $\Delta\Phi$ is the phase-difference $\Delta\Phi(t) = \Phi_\sig(t) - \Phi(t)$,
and where we have used the fact that $\int\exp[i2\pi(f+f_\sig)t]dt\approx0$. 
We see from \eref{eq:16} that $|X|$ will have a global
maximum of $|X| = 1$ if the phase-parameters of the template are perfectly
matched to the signal, i.e., $\Delta\Phi=0$, while $|X|$ will decrease
very rapidly for increasing phase-mismatches.  

The central aim of the following investigation is to determine the
regions in parameter-space for which the detection statistic $|X|$ is
of order unity for a given signal $\{\vn_\sig,\,f^{(k)}_\sig\}$.

\section{Analytic approximation for $|X|^2$}

In order to simplify the following discussion, we restrict ourselves
in this study to pulsar-signals with negligible spindown over the
observation period $T$, so we assume $f^{(k)}\approx0$ for $k>0$. 
Using \eref{eq:5}, we can therefore write the phase-mismatch
$\Delta\Phi(t)$ as
\begin{equation}
  \label{eq:18}
  \Delta\Phi(t; \vn_\sig, f_\sig; \vn, f) = 2\pi\left[ \Delta f\,t 
    + {\vr(t)\over c} \cdot \vA \right]\,,
\end{equation}
where we defined 
\begin{equation}
  \label{eq:23}
  \Delta f\equiv f_\sig - f\,,\qaq \vA\equiv f_\sig \vn_\sig - f \vn\,.
\end{equation}
We can decompose the motion of the detector $\vr(t)$ into its orbital-
and spin-component, namely, $\vr(t) = \vr_\orb(t) + \vr_\spin(t)$,
where $\vr_\orb$ is the vector from the SSB to the centre of the earth,
and $\vr_\spin$ is the vector from there to the detector.
Correspondingly we have
\begin{equation}
  \label{eq:19}
  \Delta\Phi(t) = 2\pi \Delta f \, t + \Delta\Phi_\orb(t) + \Delta\Phi_\spin(t)\,,
\end{equation}
where 
\begin{equation}
  \label{eq:35}
\Delta\Phi_\orb = {2\pi \over c} \,\vr_\orb(t)\cdot\vA\,,\qaq
\Delta\Phi_\spin = {2\pi \over c} \,\vr_\spin(t)\cdot\vA\,.
\end{equation}

\subsection{Spin motion of the earth}

Except for very short observation times $T\ll1$~day, which are not
relevant for searches for continuous waves, the phase modulation $\Phi_\spin$
due to the spin-motion of the earth is oscillatory, and can therefore
not be treated using a Taylor-expansion. However, following an
approach first used in~\cite{jotania96,valluri02}, it is more
advantageous to use the Jacobi-Anger expansion, which allows one to expand
an oscillatory exponent in terms of the Bessel-functions $J_n(z)$, namely 
\begin{equation}
  \label{eq:20}
  e^{i z \cos\zeta} = \sum_{n=-\infty}^{\infty} i^n \, J_n(z)\,e^{i n \zeta}\,.
\end{equation}
In order to be able to use this, we write the detector motion $\vr_\spin(t)$ as   
\begin{equation}
  \label{eq:8}
  \vr_\spin(t) = \sin\lat \,\RE \,\ez + \cos\lat \,\RE \,\ew(t)\,,
\end{equation}
where $\lat$ is the latitude of the detector, $\RE$ is the radius of
the earth, $\ez$ is the unit-vector along the rotation axis and
$\ew(t)$ is the time-dependent unit-vector orthogonal to $\ez$,
pointing from the rotation axis to the detector. Using this decomposition, we
can write the phase-mismatch due to the spin-motion of the earth as 
\begin{equation}
  \label{eq:9}
  \Delta\Phi_\spin(t) = \Delta\Phi_z + \KA \, \cos\varphi(t)\,,
\end{equation}
where we defined 
\begin{equation}
  \label{eq:10}
  \Delta\Phi_z \equiv 2\pi{\RE\over c}\sin\lat \, A_z\,,\quad
  \KA \equiv 2\pi {\RE\over c}\cos\lat\,A_\perp\,,
\end{equation}
and 
\begin{equation}
  \label{eq:11}
  \varphi(t) = \varphi_A + \Om_\spin\,t\,,\quad\textrm{with}\quad
  \varphi_A = \measuredangle\left(\vA_\perp, \ew(0)\right)\,,
\end{equation}
and where $A_z$ and $\vA_\perp$ are the parallel and orthogonal
projections of $\vA$ with respect to the rotation axis $\ez$.
Using this decomposition, we can apply the Jacobi-Anger expansion
\eref{eq:20} as follows:
\begin{equation}
  \label{eq:12}
  e^{i\Delta\Phi_\spin(t)} = e^{i\Delta\Phi_z} \sum_{n=-\infty}^\infty
  i^n \,e^{i n \varphi_A}\, J_n(\KA)\,e^{i \,n \,\Om_\spin\, t}\,.
\end{equation}
Substituting this together with \eref{eq:19} into the
matched-filtering amplitude \eref{eq:16}, we obtain
\begin{equation}
  \label{eq:15}
  X = e^{i\Delta\Phi_z} \sum_{n=-\infty}^\infty
  i^n \,e^{i n \varphi_A}\, J_n(\KA)\, Y_n\,,
\end{equation}
where 
\begin{equation}
  \label{eq:21}
  Y_n = {1\over T}\int \exp \left[ i \left( 2\pi\Delta f\,t +
  \Delta\Phi_\orb(t) + n \,\Om_\spin\, t\right)\right] \, d t \,.  
\end{equation}

\subsection{Orbital motion around the sun}

Contrary to the spin-motion, the orbital motion induces a slowly varying
and non-oscillatory phase-correction for observation-times $T\ll
1$~year, so we can expand the orbital phase-mismatch
$\Delta\Phi_\orb(t)$ in terms of the small quantity  $\varepsilon \equiv
\Omega_\orb T \ll 1$, which leads to 
\begin{equation}
  \label{eq:22}
  Y_n = e^{i\Delta\Phi_\orb(0)} {1\over T}\int 
  \exp\left[i \left( a_1\,t + \sum_{k=2} a_k t^k \right)\right]\,d t\,,
\end{equation}
with the coefficients
\begin{equation}
  \label{eq:33}
  \eqalign{
    {a_1 \over 2\pi} &= \Delta f + \vb_0 \cdot \vA + n\,f_\spin\,,\\
    {a_k \over 2\pi} &= {1\over k!} \,\vb_0^{\,(k-1)}\cdot\vA\,,\qquad{(k>1)}\,,
  }
\end{equation}
where we defined $f_\spin\equiv\Om_\spin/2\pi$, and where
$\vb_0 \equiv \vec{V}_0 / c = \dot{\vr}_\orb(0)/c$ denotes the orbital
velocity at the start of the observation $t=0$.  

Note that the integrals $Y_n$ will be strongly peaked as a function of
the index $n$, namely by considering
\begin{equation}
  \label{eq:24}
  |Y_n| \approx \left| {1\over T}\int e^{i a_1 t}\,d t \right|
\sim \sinc\left[ a_1 {T \over2}\right]\,,
\end{equation}
we see that $Y_n$ will have a maximum for the index $n^*$ closest to
the point $a_1=0$, i.e., 
\begin{equation}
  \label{eq:25}
  n^* = - \round\left[ {\Delta f + \vb_0 \cdot \vA \over f_\spin}
  \right]\,.  
\end{equation}
As a first approximation we therefore only keep the dominant term
$n^*$ in the Jacobi-Anger expansion, i.e., we set $Y_n \approx
Y_{n*}\,\delta_{n,n*}$, and so we obtain for the matched-filtering
amplitude 
\begin{equation}
  \label{eq:29}
  X \approx e^{i\Delta\Phi_z}\, i^{n*} \,J_{n*}(\KA) \, Y_{n*}\,.
\end{equation}
We are interested in the regions of parameter-space where
$|X|\sim\O(1)$. In these regions we can assume the exponent in
\eref{eq:22} to be small, and therefore, after Taylor-expanding and
integrating (\ref{eq:22}), we can obtain the approximate expression
\begin{equation}
  \label{eq:30}
  |X|^2 \approx J_{n*}^2 (\KA) \left[ 1 - {\hat{a}_1^2 \over 12}\,T^2 
    - {\hat{a}_1 a_2 \over 6}\, T^3 + \O(T^4)  \right] \equiv |X_1|^2\,,
\end{equation}
where $\hat{a}_1 \equiv a_1(n^*)$.

\section{The `maximum structure' of $|X|$}
\label{sec:maximum-structure-x}

\begin{figure}[htbp]
  \centering
  \includegraphics[width=0.7\textwidth,clip]{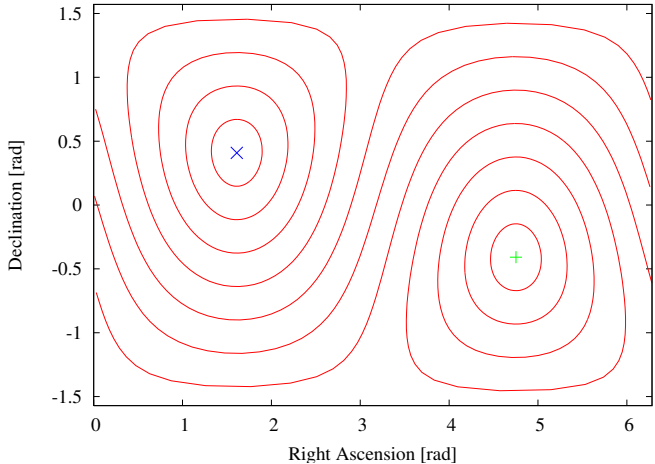}
  %% produced in octave using: plotCircleFamily (betaS2_orb, 11, 80)
  \caption{Family of circles described by \eref{eq:32} for a
    $T=10$~hours observation starting at GPS-time $t_0 = 732490940$~s.
    The direction of $\vb_1$ is marked by '$\times$', while '$+$'
    indicates the direction $-\vb_1$.  \label{fig:circles1}}
\end{figure}

On can see from its definition \eref{eq:16} that $|X|$ has a global
maximum of $|X|=1$ in the perfectly matched case ($f=f_\sig$ 
and $\vn = \vn_\sig$), while it will be nearly zero almost
everywhere else, because generally $\Delta\Phi$ will not be small. 
Guided by the observed structure of the detection statistic (see
\sref{sec:comp-fully-numer}, for example), we focus on the points
in parameter space where $|X|$ (or $\F$) has a local maximum in
target-frequency $f$ for a fixed sky-position $\vn$, i.e.,  
\begin{equation}
  \label{eq:34}
  \left. {\partial |X|^2 \over \partial f} \right|_{\vn} = 0\,.
\end{equation}
In order to understand the qualitative features of these local
maxima, we now also neglect the spin motion of the earth
(setting $n^*=0$ and $K_A=0$), and using \eref{eq:30} and
\eref{eq:33} we obtain the following condition for local maxima (in
the sense of (\ref{eq:34})): 
\begin{equation}
  \label{eq:36}
  a_1 a_1'  + a_2 \,T a_1' + a_1 \,T a_2 ' = 0\,,
\end{equation}
where we write $a_k' \equiv \partial_f a_k(f,\vn)$.
We can solve this order by order in $T$ (strictly speaking, in the
small parameter $\varepsilon= \Om_\orb T$), and to zeroth order we find
\begin{equation}
  \label{eq:38}
  \O(T^0):\quad a_1 \,a_1' = 0 \,\quad\implies\quad 
  \Delta f + \vb_0 \cdot \vA = 0\,,
\end{equation}
as $a_1' = \O(1)$. For the next order we therefore solve
\eref{eq:36} with the ansatz $a_1 = \lambda\,T$ (with $\lambda=\O(1)$),
so we find 
\begin{equation}
  \label{eq:39}
  \O(T):\quad a_1 + a_2 T = 0\,\quad\implies\quad \Delta f + \vb_1 \cdot \vA = 0\,,
\end{equation}
where $\vb_1 \equiv \vb_0 + \dot{\vb}_0\,{T\over2}$ is the first order
approximation of the orbital velocity at the midpoint of the
observation time. This scheme can be extended to arbitrary order in $T$,
but for the qualitative discussion here we restrict ourselves to first order.
The condition \eref{eq:39} can be expressed more explicitly as
\begin{equation}
  \label{eq:32}
  f\left(1 + \vb_1 \cdot\vn\right) = f_\sig\left( 1 + \vb_1 \cdot\vn_\sig\right)\,.
\end{equation}
For a given signal $\{\vn_\sig\,,f_\sig\}$, this equation describes the
(approximate) location of the local maxima of $|X|$ as a function of
$f$ for fixed $\vn$. Conversely this can be regarded at fixed
frequency $f$, describing the sky-locations $\vn$ in which the maximum
occurs at this particular frequency $f$. From the above equation one 
can easily see that these points $\vn$ describe a circle in the sky,
which has its centre in $\vb_1$, and an opening angle determined by the
signal-location and the target frequency.  
In the case $f = f_\sig$, we see that $\vn=\vn_\sig$ is one of
the possible solutions of \eref{eq:32}, and so the circle will pass
through the signal-location in the sky. If $f < f_\sig$, then we must
have $\vb_1\cdot\vn > \vb_1\cdot\vn_\sig$, i.e.,  the opening angle of
the circle has to be larger than in the case of a perfectly matched
target-frequency, and vice-versa for the case of $f > f_\sig$.
For a 10-hour integration starting at GPS-time $t_0 = 732490940$~s,
the orbital velocity at the midpoint is 
\mbox{$\vb_1= [ 0.36, -9.13, -3.96 ]\times 10^{-5}$}, and 
the corresponding family of circles is shown in \fref{fig:circles1}.  
This equation also determines the upper limit on the
frequency-mismatch $\Delta f$ for which such a local maximum is 
possible, namely, by rewriting \eref{eq:32} as 
\begin{equation}
  \label{eq:37}
  \left| \Delta f \over f_\sig \right| 
  = { \left| \vb_1 \cdot \Delta\vn \right| \over \left| 1 + \vb_1 \cdot \vn\right|}
  \le {2 \beta_1 \over 1 - \beta_1 } = 2\beta_1 + \O(\beta_1^2)\,,
\end{equation}
we see that the maximal relative frequency-mismatch is bounded by
\emph{twice} the Doppler-shift due to the orbital velocity. Using the
value $V_\orb \approx 3\times10^{4}$~m~s$^{-1}$, this corresponds to a
``Doppler-window'' of $\Delta f/f \sim \pm 2\times10^{-4}$.

We note another interesting property of the approximate expression
\eref{eq:30} for the matched-filtering amplitude $|X|$: namely,
from the above discussion we know that the local maxima of $|X|$
occur on or close to the circles described by \eref{eq:32}, which
are characterised by $a_1\approx 0$. Therefore we see from \eref{eq:30}
that the amplitude of $|X_1|$ along these circles will be dominated by
the factor $J_{n*}(K_A)$, where $K_A$ is proportional to
$A_\perp$, the projection of $\vA$ into the equatorial plane. This
factor is therefore symmetrical in the target sky-position $\vn$ with
respect to the equatorial plane, a feature which can indeed be seen in
the plot of \eref{eq:30} shown in \fref{fig:Xsq}(b).
Interestingly, this symmetry \emph{can} still be present in the
fully numerical $\F$-statistic, as seen, for example, in
\fref{fig:FstatSignalOnly10hours}(b), while the effects of
spin-motion and antenna-patterns can also mask this feature depending
on the parameters, as seen in \fref{fig:FstatSignalOnly10hours}(a). 

\section{Comparison to fully numerical results for the $\F$-statistic}
\label{sec:comp-fully-numer}

We now turn to comparing these analytic results to the actual
parameter-space structure of the fully numerical $\F$-statistic
including all effects of amplitude-modulation and using an
ephemeris-based description of the detector-motion.
As an example, we consider a 10-hour observation time starting at
GPS-time $t_0 = 732490940$~s, and we use a fake pulsar-signal at
$f_\sig=100$~Hz and sky-position $(\alpha_\sig,\;\delta_\sig)=(2,1)$,
for two different choices of amplitude-parameters shown in \tref{tab:pulsars}.
In the following the detector-location is always chosen to be the
LIGO~Livingston~Observatory (LLO), and we only consider pure signals
without noise. 
\begin{table}[htb]
  \caption{\label{tab:pulsars}Amplitude- and phase-parameters of two fake pulsar-signals.}
  \begin{indented}
    \item[] \begin{tabular}{@{}lcccccc}
        \br
        Model     & $f$ [Hz] & $(\alpha,\delta)$ [rad] & $A_+$ & $A_\cross$ & $\Psi$ & $\Phi_0$ \\
        \mr
        I  & $100$ & $(2.0,\; 1.0)$ & $1.0$ & $0.5$ & $1.0$ & $2.0$  \\
        II & $100$ & $(2.0,\; 1.0)$ & $1.0$ & $0.0$ & $0.0$ & $0.0$ \\
        \br
      \end{tabular}
    \end{indented}
\end{table}
The fake pulsar-signal is generated using 
\texttt{makefakedata\_v4}, and the $\F$-statistic~\cite{jks98:_data} is 
calculated using \texttt{ComputeFStatistic}\footnote{These codes are
  found in \texttt{LALApps}\cite{lalapps}, under
  \texttt{src/pulsar/Injections} and \texttt{src/pulsar/FDS\_isolated}.}.

\begin{figure}[htbp]
  \centering
  \includegraphics[width=0.8\textwidth,clip]{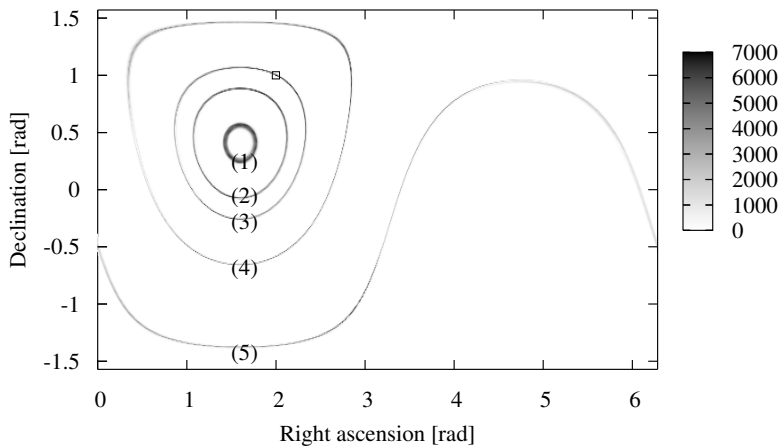}
  \caption{Plot of the $\F$-statistic over the sky superposed for
    different target frequencies $f$: (1) $f = 100.002\;\Hz$, 
    (2) $f = 100.001\;\Hz$, 
    (3) $f = f_\sig=100\;\Hz$, (4) $f = 99.997\;\Hz$ and (5)
    $f=99.99\;\Hz$.
    The parameters of the injected signal correspond to model~I in
    \tref{tab:pulsars}, and the small square indicates the sky-position
    of the signal.}
  \label{fig:circlesF}
\end{figure}

We observe the following structure of the $\F$-statistic over the
parameter space $\{\alpha, \delta, f\}$: searching over the sky at a
perfectly matched target-frequency $f = f_\sig$, the 
$\F$-statistic has a maximum at the signal sky-position $(\alpha_\sig,
\delta_\sig)$, as expected, but remains of the same order on a very
narrow, complete circle over the sky (obviously including the
signal-position), while it is practically zero everywhere else. 
For mismatched target-frequencies $f$ (within the Doppler-window of $f_\sig$), 
there is a different but concentric circle of $\F$-values comparable to
the maximum, with a larger radius for $f < f_\sig$, and a smaller
radius for $f > f_\sig$, as shown in \fref{fig:circlesF}.  
This agrees qualitatively and quantitatively surprisingly well with
the circle-structure predicted solely based on the effect of the
orbital Doppler-shift, i.e., \eref{eq:32} and \fref{fig:circles1}.

\begin{figure}[htbp]
  \centering
  \mbox{
    \parbox[t]{0.5\textwidth}{
      (a)\\
      \hspace*{-0.8cm}
      \includegraphics[width=0.55\textwidth,clip]{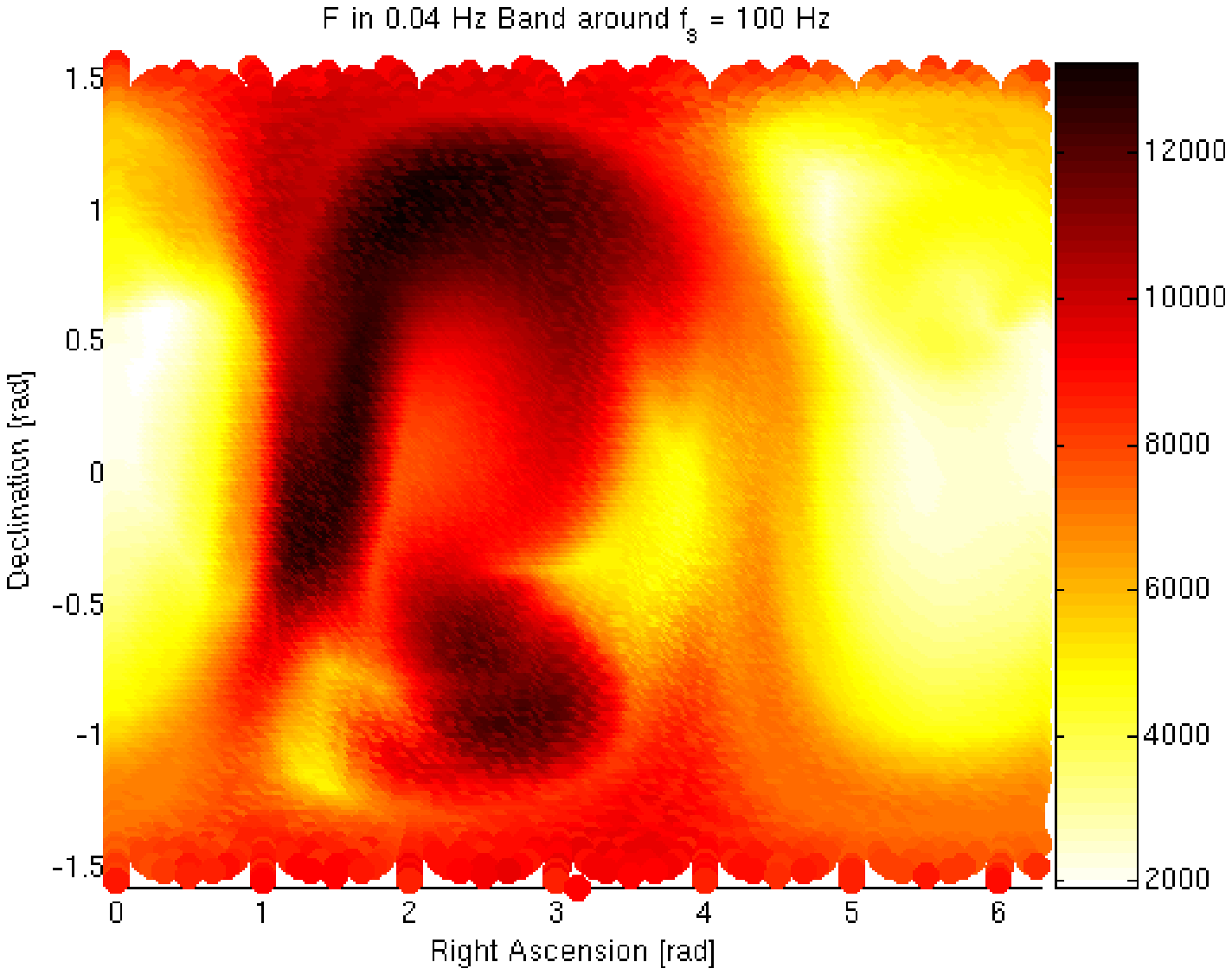}
    }
    \parbox[t]{0.5\textwidth}{
      (b)\\
      \includegraphics[width=0.55\textwidth,clip]{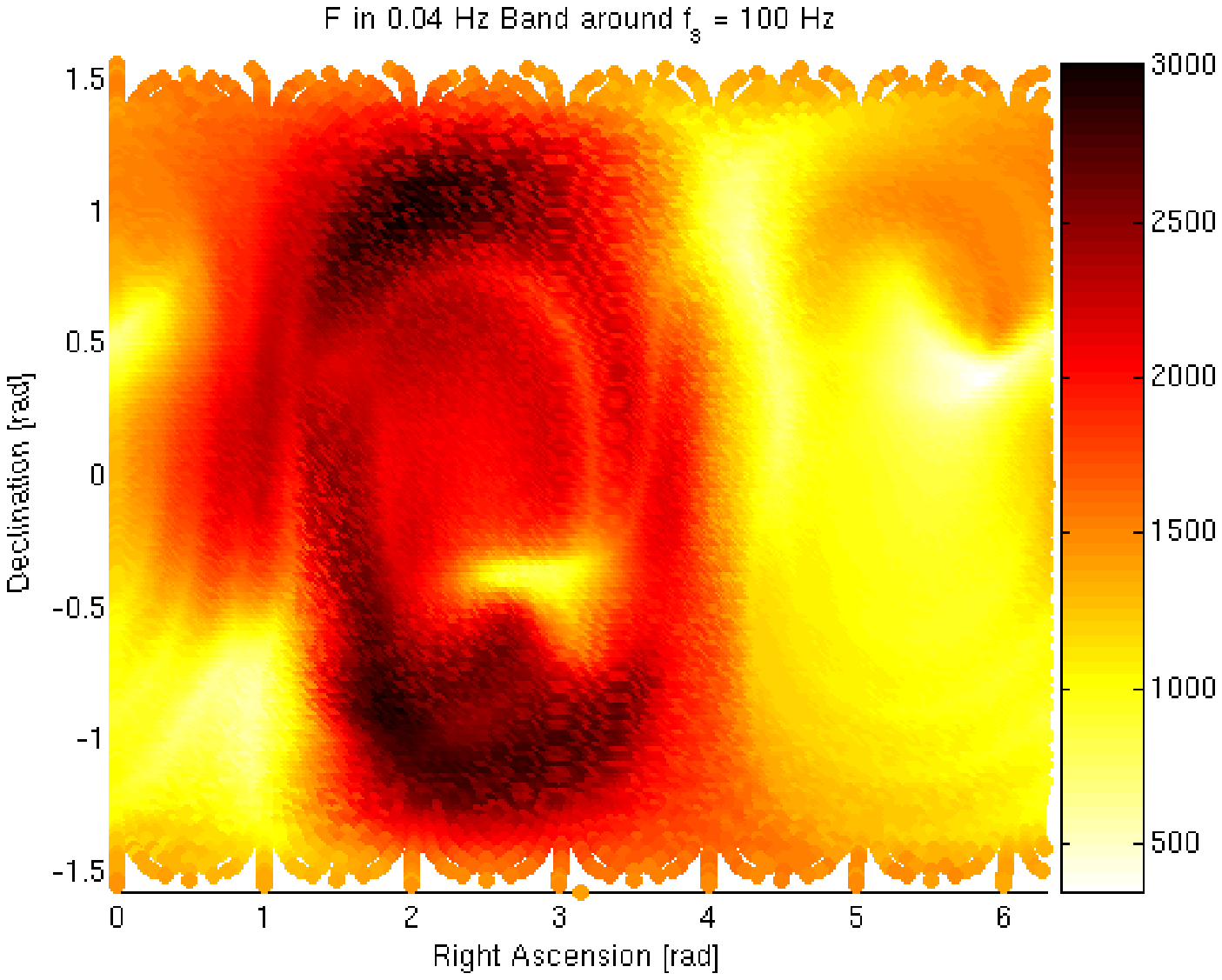}
    }
  }
  \caption{
    Projected $\F$-statistic over the sky in the frequency range 
    $f\in (1\pm2\times10^{-4})\,f_\sig$ for (a) pulsar-signal~I and 
    (b) signal~II as defined in \tref{tab:pulsars}.}
  \label{fig:FstatSignalOnly10hours}
\end{figure}

In the next step we run an all-sky search over the whole Doppler-window
\mbox{$f\in (1\pm2\times10^{-4})\,f_\sig$}, corresponding to a dense
superposition of these circles for the different target-frequencies. The
resulting $\F$-statistic projected on the sky is shown in
\fref{fig:FstatSignalOnly10hours}(a) and (b) for pulsar-signals I and II,
respectively.  
From these projected plots we see more clearly that the variation of
the amplitude of the $\F$-statistic maxima is more sensitive to the
polarisation-parameters of the signal, due to effects of the
antenna-pattern. 
For example, \fref{fig:FstatSignalOnly10hours}(b) shows some of the
equatorial symmetry characteristic for the approximate expression
\eref{eq:30}, as discussed in \sref{sec:maximum-structure-x}, in
particular we see a mirror-image in $(2, -1)$ of the absolute maximum
at the signal-position $(2,1)$. For signal~I, on the other hand, 
the projected $\F$ is more asymmetrical and there is no
mirror-image of the maximum in the southern hemisphere.

Summarising these observations we can state that the \emph{locations}
of the local maxima of $\F$ are dominantly determined by the orbital
velocity of the earth, and are described very 
well\footnote{A preliminary numerical comparison of the frequency
  $f_\F$ of the maximum of $\F$ for a fixed sky-position $\vn$ and the
  theoretical prediction $f_\th$ from \eref{eq:32} yields the bound
  $|f_\F - f_\th|/f_\F < 10^{-6}$ over the whole sky.}
by the circles \eref{eq:32}, independently of detector-location
and amplitude-parameters of the signal. The spin-motion of the
detector and amplitude-modulation due to the rotating antenna-pattern,
on the other hand, are responsible for the variation of the amplitude
of $\F$ along these circles, as seen in \fref{fig:FstatSignalOnly10hours}.    

\begin{figure}[htbp]
  \centering
  \mbox{
    \parbox[t]{0.5\textwidth}{
      (a)\\
      \hspace*{-0.8cm}\includegraphics[width=0.65\textwidth,clip]{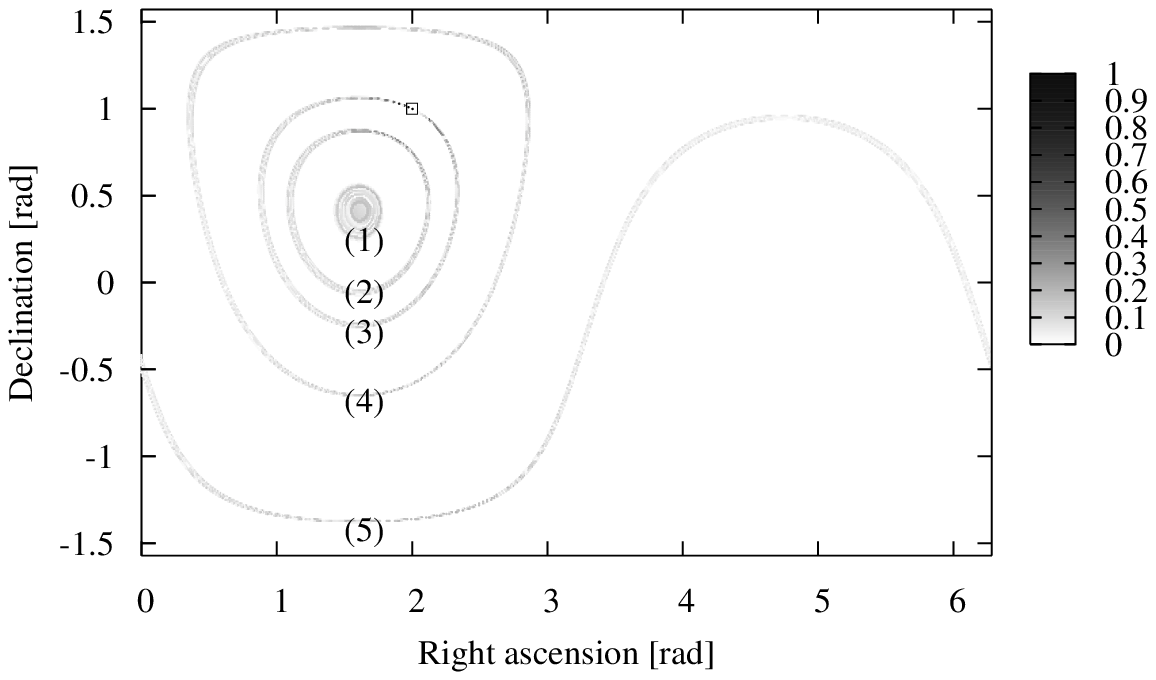}
    }
    \hspace{1.0cm}
    \parbox[t]{0.5\textwidth}{
      (b)\\
      \includegraphics[width=0.47\textwidth,clip]{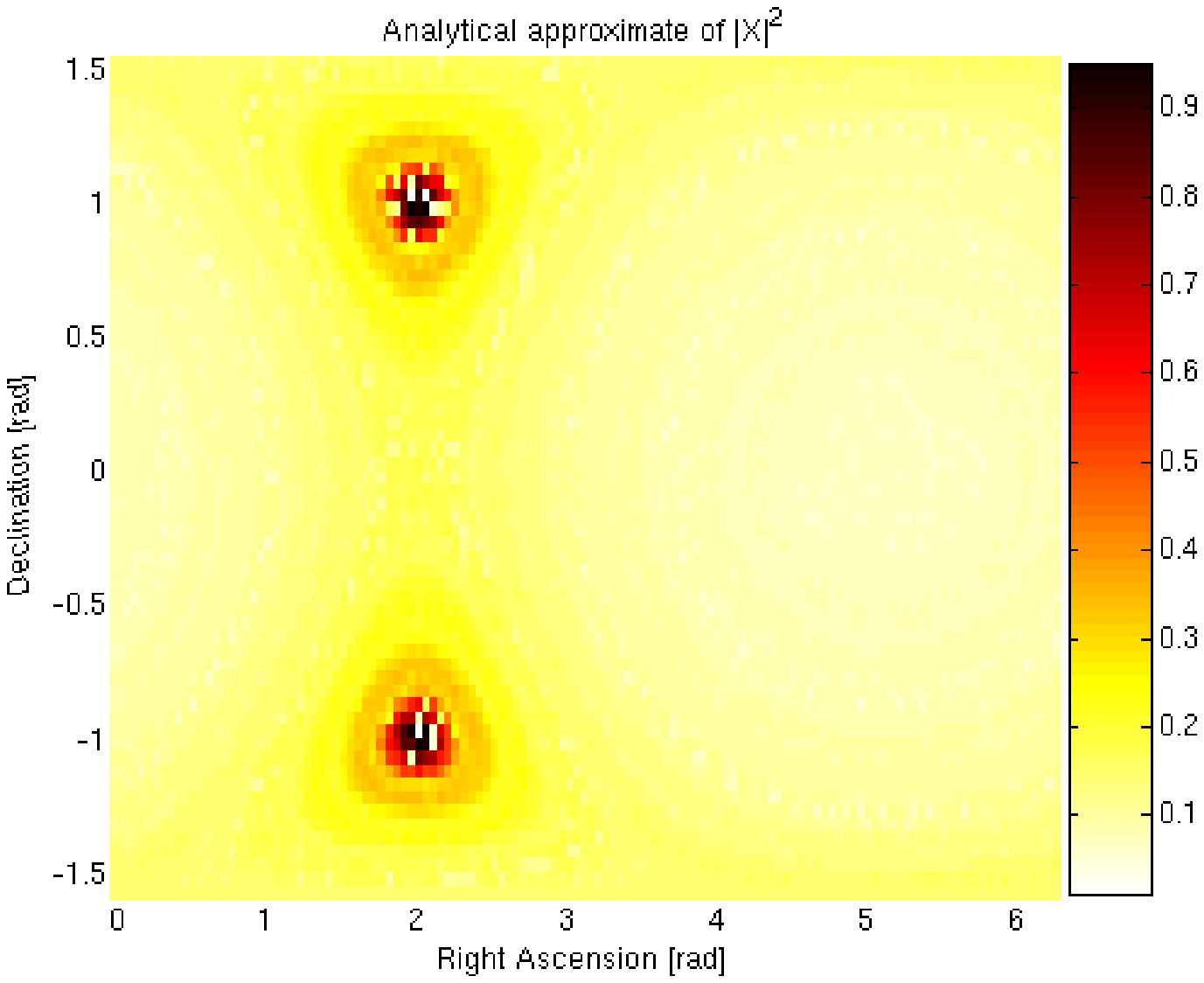}
    }
  }
  \caption{Plot of the approximate matched-filtering amplitude $|X_1|^2$ 
    over the sky for (a) different target frequencies (1) $f = 100.002\;\Hz$, 
    (2) $f = 100.001\;\Hz$, (3) $f = f_\sig=100\;\Hz$, 
    (4) $f = 99.997\;\Hz$ and (5) $f=99.99\;\Hz$, and
    (b) over the whole Doppler-window $f\in (1\pm2\times10^{-4})\,f_\sig$ 
    projected on the sky. The small square indicates the sky-position
    of the signal (model~I in \tref{tab:pulsars}). 
  }
  \label{fig:Xsq}
\end{figure}

Finally, we plot the approximate detection amplitude $|X_1|^2$ given by
\eref{eq:30} over the sky for different target frequencies, as shown
in \fref{fig:Xsq}(a), which shows some striking similarities to the
`exact' result in \fref{fig:circlesF}. 
In particular, it exhibits the feature of vanishing very rapidly
outside the narrow ``circles'' described approximately by \eref{eq:32}. 
The amplitude of $|X_1|^2$ on the circles, however, is not a good
description of the corresponding $\F$-statistic amplitude, as $|X_1|^2$
is seen to be much more strongly peaked around the signal-location and
decreases rather rapidly to about $0.1$ on most of the circles
(see \fref{fig:Xsq}(b)), while the $\F$-statistic has a similar
amplitude to the maximum over most of the circles. Note also 
that \fref{fig:Xsq}(b) illustrates the equatorial symmetry of
$|X_1|^2$ as discussed at the end of \sref{sec:maximum-structure-x}.

\section{Discussion}

In a practical search for continuous gravitational waves we will
usually be interested in the highest values (``candidates'') of $\F$
over a chosen volume of the parameter-space.
From the results presented here we expect that a real signal would
generate candidates nearly everywhere on the sky within the
Doppler-window $\pm 2\times10^{-4}$ of the signal-frequency $f_\sig$. 
All these candidates, however, will satisfy \eref{eq:32}
with an (approximately) identical value on the right-hand side, and so
the quantity $\kappa \equiv f(1 + \vb_1\cdot\vn)$ is an (approximate)
invariant of candidates caused by the same signal. This could be a
very useful criterion for classifying equivalent candidates, and also
for determining coincident candidates from different detectors.
One could further use this to construct coincidence-tests of
candidates from different observation times, as $\vb_1(t)$ is a known
function of time. 
In order for these methods to be practically useful, however, 
the errors of the different approximations need to be quantified, 
and we need to test the robustness of the suggested coincidence criteria
in the presence of noise. 

\ack
We would like to thank Badri Krishnan, Maria Alessandra Papa, Xavier Siemens 
and Alicia Sintes for helpful discussions.

\end{document}